\documentstyle{article}
\begin{document}
\rightline{Feb 1994}
\rightline{McGill/94-11}
\vskip 2cm
\begin{center}
\begin{large}
{\bf A parity invariant $SU(3)_c \otimes SU(3)_L \otimes U(1)$ model}
\vskip 2cm
\end{large}
R. Foot\footnote{email: Foot@hep.physics.mcgill.ca}\\
Department of Physics,
McGill University,\\
3600 University Street,
Montreal, Quebec, Canada H3A 2T8.\\

\end{center}
\vskip 1cm
ABSTRACT
\vskip .5cm
\noindent
We examine a $SU(3)_c \otimes SU(3)_{L}\otimes U(1)$ gauge model
which has a parity
symmetric Lagrangian. The parity symmetry has the novel
feature that it interchanges the gluons with the $SU(3)_L$
gauge bosons (which contain the ordinary $SU(2)_L$ weak gauge
bosons). We show that the model reduces to the
standard model at low energies and also predicts new
physics in the form of exotic fermions.
\newpage

The standard model has been rather successful in recent times.
With the present lack of experimental guidence it is difficult to know
what lies beyond the standard model. One possibility is that
nature respects some type of symmetry not present in the standard
model. In particular, discrete space-time symmetries (parity, time
reversal and charge conjugation) may be symmetries of the Lagrangian
describing nature, but which are not manifested in the standard
model Lagrangian because it may be incomplete. In this  sense,
discrete space-time symmetries may have a {\it fundamental} role to
play in nature even though they are not
symmetries of the Standard model Lagrangian.
It is an interesting possibility that nature is described by a Lagrangian
which is invariant under one or all of the discrete space-time symmetries.
One well known example is the usual left-right symmetric model with
gauge group $SU(3)_c \otimes SU(2)_L \otimes SU(2)_R \otimes U(1)$
and is invariant under a parity symmetry [1]. This parity symmetry
interchanges the $SU(2)_L$ gauge bosons with the $SU(2)_R$
gauge bosons and is assumed to be spontaneously broken by the vacuum.
The usual left-right symmetric model is not the only possible
model which allows parity to be a symmetry of the Lagrangian.
For example, the quark-lepton symmetry model has gauge group
$SU(3)_q \otimes SU(3)_{l} \otimes SU(2)_L \otimes U(1)$ and has a
discrete symmetry which interchanges quarks and leptons [2]. This
discrete symmetry can also be interpreted as parity (which is assumed
to be spontaneously broken by the vacuum) and is thus another
realization of a gauge model with left-right symmetry spontaneously
broken [3].

The purpose of this letter is to point out that there is another
type of model with left-right symmetry which is spontaneously broken.
The idea is to build a model in which parity interchanges the gluons
with the weak gauge bosons ($W_L$'s) (rather than $W_L's$ interchanging
with $W_R's$ as is the case in the usual left-right symmetric model).
In-order for parity to interchange gluons with $W_L$'s we need
to enlarge the standard model gauge symmetry to:
$$SU(3)_c \otimes SU(3)_L \otimes U(1)_{y'} \eqno (1)$$
We then need to construct the fermions so that they are symmetric when
the two $SU(3)$'s are interchanged. The model must also reduce to the standard
model at some scale if it is to describe the real world.
This can be achieved with the following fermion representations, (under
the gauge symmetry eq.(1)):
$$\begin{array}{ccc}
f_L^a \sim (1, 3, -1/3),& \
f{'}_L^a \sim (1, 3, -1/3),& \  F_L^a \sim (1, 3, 2/3),\\
 d_R^a \sim (3, 1, -1/3),& \
d{'}_R^a \sim (3, 1, -1/3),&\ u_R^a \sim (3, 1, 2/3),\\
Q_L^a \sim (3, \bar 3, 0),& & \end{array} \eqno (2)$$
where $a = 1, 2, 3,$ is the generation index [4].
It is straightforward to check that with the above choice
of gauge quantum numbers, all of the gauge anomalies cancel per
generation.
We now introduce the discrete parity operation which we
define below:
$$ x \rightarrow -x,\  t \rightarrow t, $$
$$ G^a_{c\mu} \leftrightarrow W^{a\mu}_L,\  B_{\mu} \leftrightarrow B^{\mu},$$
$$f_L \leftrightarrow \gamma_0 d_R,\ f'_L \leftrightarrow \gamma_0 d'_R,
\  u_R \leftrightarrow \gamma_0 F_L,
\ Q_L \leftrightarrow \gamma_0 (Q_L)^c, \eqno(3)$$
where $G^a_{c\mu},\ W^a_{L\mu},\ B_{\mu}$ are the gauge bosons of $SU(3)_c,\
SU(3)_L$ and $U(1)_{y'}$ respectively. It is straightforward to check that
the gauge covariant kinetic terms of the fermions and the gauge bosons
are invariant with respect to the parity symmetry (eq.(3)) provided
that the bare gauge coupling constants of $SU(3)_c$ and $SU(3)_L$ are
equal.

Gauge symmetry breaking to the standard model can be achieved by introducing
a Higgs field, $\chi_1 \sim (1, \bar 3, 1/3)$. The field $\chi_1$ couples to
the fermions of the model as follows:
$${\cal L}^{\chi_1} = \lambda_1 \bar Q_L \chi_1 d'_R +
\lambda_2 \bar f'_L \chi_1 (F_L)^c + H.c., \eqno (4)$$
where we have suppressed the generation indicies.
Provided that $\chi_1$ develops the vacuum expectation value (VEV):
$$\langle \chi_1 \rangle
= \left(\begin{array}{c}
0\\
0\\
w\end{array}\right), \eqno (5)$$
then the gauge symmetry breaks down to the Standard model:
$$\begin{array}{c}
SU(3)_c \otimes SU(3)_L \otimes U(1)_{y'}\\
\downarrow \langle \chi_1 \rangle\\
SU(3)_c \otimes SU(2)_L \otimes U(1)_Y
\end{array}\eqno (6)$$
where $Y = 2y' - \lambda_8/\sqrt{3}$ is the combination of $y'$ and
$\lambda_8$ which is left unbroken by $\langle \chi_1 \rangle$ ($\lambda_8 =
diag(1, 1, -2)/\sqrt{3}$).
{}From eq.(2) it is easy to see that each generation contains 27 Weyl
fermions in this model. One can identify 15 of these Weyl fermions
as the standard quarks and leptons with the correct gauge quantum
numbers under the $SU(3)_c \otimes SU(2)_L \otimes U(1)_Y$ subgroup left
unbroken by the VEV $\langle \chi_1 \rangle$.
In particular, note that $Y (= 2y' - \lambda_8/\sqrt{3})$,
is numerically identical to the
standard model hypercharge assignments.
In addition to these standard 15 Weyl fermions, there are 12 exotic
fermion fields per generation. Under the $SU(3)_c \otimes SU(2)_L
\otimes U(1)_Y$ subgroup, these 12 exotic fermions have the following gauge
quantum numbers: $d'_{L,R} \sim (3, 1, -2/3), \ X'_{L,R} = (V_{L,R},
E_{L,R})^T \sim
(1, 2, -1), \ \nu_R, \nu^{'}_R \sim (1, 1, 0)$. We identify an exotic
$SU(2)_L$ singlet charged -1/3 quark, and a leptonic (i.e.
color singlet) vector-like $SU(2)_L$ doublet of leptons, together with
two singlet Weyl neutrinos. Observe that
the exotic charged $-1/3$ quarks gain masses and the exotic lepton
doublet gain
masses from the Yukawa Lagrangian terms in eq.(4) when
$\chi_1$ develops the VEV $\langle \chi_1 \rangle$ (eq.(5)).
In order to fascillitate understanding we can write the
$SU(3)_L$ degree's of freedom explicitly:
$$\begin{array}{ccc}
f_L = \left(\begin{array}{c}
\nu_L\\
e_L\\
(\nu_R)^c\end{array}\right), &\  f'_L = \left(\begin{array}{c}
V_L\\
E_L\\
(\nu'_R)^c\end{array}\right),& \  F_L = \left(\begin{array}{c}
(E_R)^c \\
(V_R)^c\\
(e_R)^c\end{array}\right),\\
Q_L = \left(\begin{array}{c}
d_L\\
u_L\\
d'_L\end{array}\right), &\ d_R,\  u_R, \ d'_R& \end{array}
\eqno (7)$$

In order to preserve the parity symmetry in the Yukawa sector,
we need to introduce a scalar
mulitplet $\chi_2$ in addition to $\chi_1$. The scalar
$\chi_2$ has the ``mirror''
gauge quantum numbers i.e. $\chi_2 \sim (\bar 3, 1, 1/3)$ and couples
to the fermions as follows:
$${\cal L}^{\chi_2} = \lambda_1 \overline{(Q_L)^c} \chi_2 f'_L +
\lambda_2 \bar d'_R \chi_2 (u_R)^c + H.c.
 \eqno (8)$$
Note that under parity, ${\cal L}^{\chi_1} \leftrightarrow {\cal L}^{\chi_2}$,
and the total Lagrangian remains invariant. Also note that we must require
$\chi_2$ to have zero VEV if we want to keep $SU(3)_c$ unbroken.

Ordinary electroweak symmetry breaking and fermion mass generation can be
achieved by introducing two $SU(3)_L$ triplets $\rho_1, \ \eta_1$:
$$\begin{array}{ll}
{\cal L}^{\eta_1, \rho_1} =& \lambda_1^{'}\bar Q_L d_R \rho_1 +
\lambda_2^{'} \bar Q_L d'_R \rho_1 + \lambda_3^{'}\bar Q_L u_R \eta_1 +
\lambda_4^{'} \bar f_L (F_L)^c \rho_1 +
\lambda_5^{'} \bar f'_L (F_L)^c \rho_1\\
& + \lambda_6^{'} \bar f_L \eta_1 (f_L)^c
+ \lambda_7^{'} \bar f_L \eta_1 (f'_L)^c +
+ \lambda_8^{'} \bar f'_L \eta_1 (f'_L)^c + H.c. \end{array}\eqno (9)$$
where the generation indices have been suppressed and
$\rho_1 \sim (1, \bar 3, 1/3),\ \eta_1 \sim (1, \bar 3, -2/3)$. Note that
we require that $\rho_1, \eta_1$ develop the VEVs:
$$\langle \rho_1 \rangle
= \left(\begin{array}{c}
u\\
0\\
0\end{array}\right), \ \langle \eta_1 \rangle
= \left(\begin{array}{c}
0\\
v\\
0\end{array}\right). \eqno (10a,b)$$

These VEVs also give masses to the usual W, Z bosons, and the standard model
intermediate gauge symmetry is
broken in the usual way to $SU(3)_c \otimes U(1)_Q$:
$$\begin{array}{c}
SU(3)_c \otimes SU(3)_L \otimes U(1)_{y'}\\
\downarrow \langle \chi_1 \rangle\\
SU(3)_c \otimes SU(2)_L \otimes U(1)_Y\\
\downarrow \langle \rho_1 \rangle, \langle \eta_1 \rangle\\
SU(3)_c \otimes U(1)_Q
\end{array}\eqno (11)$$
where $Q = I_3 + Y/2$, and is of course just the usual electric charge
operator.

In order to maintain the discrete symmetry, the mirror partners $\rho_2,
\eta_2$
also need to be introduced and as in the case of  $\chi_2$, they must
gain zero VEV. It is easy to deduce their coupling to fermions, which
is of course dictated by the requirement that the discrete symmetry
be a symmetry of the Lagrangian:
$$\begin{array}{l}
{\cal L}^{\eta_2, \rho_2} = \lambda_1^{'}\overline{(Q_L)^c} f_L \rho_2 +
\lambda_2^{'} \overline{(Q_L)^c} f'_L \rho_2 +
\lambda_3^{'}\overline{(Q_L)^c} F_L \eta_2 +
\lambda_4^{'} \bar d_R (u_R)^c \rho_2 \\
+
\lambda_5^{'} \bar d'_R (u_R)^c \rho_2 + \lambda_6^{'} \bar d_R \eta_2 (d_R)^c
+ \lambda_7^{'} \bar d_R \eta_2 (d'_R)^c +
+ \lambda_8^{'} \bar d'_R \eta_2 (d'_R)^c + H.c. \end{array}\eqno (12)$$

The existence of the discrete symmetry means that the bare $SU(3)_c$ and
$SU(3)_L$ coupling constants must be equal to each other.
This unfortunately tends to imply that $SU(3)_L$ symmetry breaking must
occur at a very high energy scale (near $10^{15}$ GeV). This means that with
the simplest symmetry breaking scenario given here, it won't be possible
to detect any of the new gauge bosons predicted by the model. However,
it may be possible to detect the exotic fermions or
the effects of the mixing. The exotic charged
$-1/3$ quarks can mix with the ordinary $d$ quark, which will
imply a small loss of GIM cancellation (since the
exotic $d'$ quark has non-canonical $SU(2)_L \otimes U(1)_Y$
gauge quantum numbers). The phenomenology of this has
already been studied in the liturature [5].

The model can also incorporate neutrino masses. This may be important
in the light of the solar neutrino problem [6].
One interesting point is that if
$\lambda_6^{'} \gg \lambda_7^{'}$ (in eq.(9)),
then two neutrino masses are nearly degenerate (in the limit that
$\lambda_7^{'} = 0$, two of the ordinary weakly interacting neutrinos
gain degerate Dirac masses while one remains massless [7].
In the MSW solution of the solar neutrino problem, the neutrino
mass difference
needs to be very small ($\sim 10^{-4}$ eV). This can be achieved
naturally if either one
neutrino is much heavier than the other or they are
nearly degenerate. The latter case is perhaps phenomenologically more
interesting because it means that the neutrino masses can be a lot
larger than their mass difference and thus it may be possible to detect
the neutrino masses directly in the Laboratory.

Finally note that the model has electric charge quantization classically [8].
By this, I mean that the classical structure of the Lagrangian
such as the gauge invariance of the Yukawa Lagrangians (eq.(4) and eq.(9)),
and the discrete parity symmetry imply that the $U(1)_{y'}$ charges of the
fermions {\it and} the scalars are uniquely fixed (upto an overall
normalization factor, which can be absorbed into the definition of
the coupling constant). This fact then implies that the electric
charges are also uniquely determined (up to an overall
normalization factor), and thus every electric charge ratio
is fixed in this theory.

In conclusion, we have proposed a simple gauge model which is
defined by a Lagrangian with a novel realization of parity
symmetry. The parity operation interchanges the gluons of QCD
with the weak gauge bosons of an extended $SU(3)_L$
weak interaction. We have pointed out that the gauge model reduces successfully
to the standard model at low energies. The principle phenomenological
feature of the model is the prediction of new fermions.
In particular exotic $SU(2)_L$ singlet charged -1/3 quarks and
exotic leptons together with two right-handed neutrinos per generation
are predicted. Neutrinos are expected to be massive in the model,
and the model may be relevant to the MSW solution of the solar neutrino
problem.
\newpage
\vskip .2cm
\noindent
{\bf References}

\vskip .7cm
\noindent
[1] J. C. Pati and A. Salam, Phys. Rev. D10, 1502 (1974); R. N. Mohapatra and
J. C. Pati, Phys. Rev. D11, 566, 2558 (1974);
G. Senjanovic and R. N. Mohapatra,
Phys. Rev. D12, 1502 (1974).
\vskip .4cm
\noindent
[2] R. Foot and H. Lew, Phys. Rev. D41, 3502 (1990); R. Foot, H. Lew
and R. R. Volkas, Mod. Phys. Lett. A8, 1859 (1993).

\vskip .4cm
\noindent
[3] It is also possible to define a gauge model in which
parity is a symmetry of the Lagrangian and the symmetry
is {\it not} broken by the vacuum but remains exact after symmetry
breaking. See R. Foot, H. Lew and R. R. Volkas, Phys. Lett. B272, 67 (1991);
Mod. Phys. Lett. A7, 2567 (1992).

\vskip .4cm
\noindent
[4] The model is related to the $SU(3)^3$ model (See e.g. A. de Rujula,
H. Georgi and S. L. Glashow, in Fifth
Workshop on Grand Unification, edited by K. Kang, H. Fried and P. Frampton,
(World Scientific, Singapore, 1984), however
that model does not have a $Z_2$ parity symmetric Lagrangian
(although it does have a $Z_3$ symmetry). Some of the phenomenology
of the two models is similiar, e.g. they have the same number
of exotic fermions with the same gauge quantum numbers under the
$SU(3)_c \otimes SU(2)_L \otimes U(1)_Y$ subgroup.
My model (like the $SU(3)^3$ model)
is also related to $E_6$ models, and the 27 Weyl fermions
in eq.(2) can fit into the 27 irreducible representation of $E_6$,
however, again the $E_6$ model is explicitly parity violating.
My motivation for the $SU(3)^2 \otimes U(1)$ model is that it is
a new type of model which has a parity conserving Lagrangian.
\vskip .4cm
\noindent
[5] See for example, P. Langacker and D. London, Phys. Rev. D38, 866 (1988).
\vskip .4cm
\noindent
[6] For a review of the solar neutrino problem, see for example
J. N. Bahcall, Neutrino astrophysics (CUP), 1989.
\vskip .4cm
\noindent
[7] To see
this note that the $\lambda_6^{'}$ term gives
rise to a $3 \times 3$ mass matrix which is antisymmetric in generation
space. (This becomes evident by noting that under interchange of the
two identical $f_L's$, the Lorentz contraction is antisymmetric, there is
an antisymmetric factor because of Fermi-Dirac statistics, and there is
a antisymmetry because of the $SU(3)_L$ contraction is antisymmetric.
Thus only the anti-symmetric piece in generation space is non-vannishing.)
A $3 \times 3$ antisymmetric mass matrix has two degenerate eigenvalues
and one zero eigenvalue.
\vskip .5cm
\noindent
[8] For a review of electric charge quantization in the standard model and
beyond, see R. Foot, G. C. Joshi, H. Lew and R. R. Volkas, Mod. Phys. Lett.
A5, 2721 (1990).
\end{document}